\documentclass[aps,prxquantum,reprint,nofootinbib,superscriptaddress]{revtex4-2}
\usepackage{graphicx}
\usepackage{dcolumn}
\usepackage{bm}
\usepackage{dsfont}
\usepackage{amsmath}
\usepackage{hyperref}
\usepackage{float}
\usepackage{caption}
\usepackage{natbib}
\usepackage{graphicx}
\usepackage{bbold}
\usepackage{physics}
\usepackage{mathtools}
\usepackage{MnSymbol}
\usepackage{relsize}
\usepackage{simpler-wick}
\usepackage{subfigure}
\usepackage{ragged2e}

\usepackage{pgfplots}
\pgfplotsset{compat=1.18} 
\usepackage{xcolor}
\definecolor{myblue}{HTML}{1f77b4}

\renewcommand{\i}{\mathrm{i}}
\newcommand{\e}{\mathrm{e}}
\graphicspath{ {./imgs/} }
\DeclareGraphicsExtensions{.pdf,.png}

\usepackage{natbib}
\begin{document}
\title{Line search by quantum logic spectroscopy enhanced with squeezing and statistical tests}
\author{Ivan Vybornyi}
\affiliation{Institut für Theoretische Physik, Leibniz Universität Hannover, Appelstrasse 2, 30167 Hannover, Germany}
\author{Shuying Chen}
\affiliation{QUEST Institute for Experimental Quantum Metrology, Physikalisch-Technische Bundesanstalt, Bundesallee 100, Braunschweig 38116 Germany}
\author{Lukas J. Spieß}
\affiliation{QUEST Institute for Experimental Quantum Metrology, Physikalisch-Technische Bundesanstalt, Bundesallee 100, Braunschweig 38116 Germany}
\author{Piet O. Schmidt}
\affiliation{QUEST Institute for Experimental Quantum Metrology, Physikalisch-Technische Bundesanstalt, Bundesallee 100, Braunschweig 38116 Germany}
\affiliation{Institut für Quantenoptik, Leibniz Universität Hannover, Welfengarten 1, Hannover 30167 Germany}
\author{Klemens Hammerer}
\affiliation{Institut für Theoretische Physik, Leibniz Universität Hannover, Appelstrasse 2, 30167 Hannover, Germany}

\date\today
\begin{abstract}
In quantum logic spectroscopy, internal transitions of trapped ions and molecules can be probed by measuring the motional displacement caused by an applied light field of variable frequency. This provides a solution to ``needle in a haystack'' problems, such as the search for narrow clock transitions in highly charged ions, recently discussed by S. Chen et al. (Phys. Rev. Applied $\boldsymbol{22}$, 054059). The main bottleneck is the search speed over a frequency bandwidth, which can be increased by enhancing the sensitivity of displacement detection. In this work, we explore two complementary improvements: the use of squeezed motional states\textcolor{black}{, explained using an analytical phase space model} and optimal statistical postprocessing of data using a hypothesis testing framework. We demonstrate that each method independently provides a substantial boost to search speed. Their combination effectively mitigates state preparation and measurement errors, improving the search speed by an order of magnitude and fully leveraging the quantum enhancement offered by squeezing. 
\end{abstract}

\maketitle
\section{Introduction}
High control over the motional and the electronic degrees of freedom of trapped ions makes them great sensors, \textcolor{black}{enabling} quantum technology applications and searches for new physics \cite{Bruzewicz2019,Hur2022,Dreissen2022,PhysRevA.110.042823,Cairncross2017}. At the heart of many trapped-ion quantum metrology schemes lies detection of weak motional displacements. In particular, it is a working principle of quantum logic sensing protocols used to probe the electronic structure of otherwise inaccessible ions \cite{Schmidt2005,PhysRevA.105.063709}. Recently, such a technique was demonstrated to be of great use to search for narrow optical transitions in highly-charged species \cite{cheung_finding_2025,Chen2024}. The latter are promising candidates for \textcolor{black}{a} new generation of atomic clocks \cite{Kozlov2018,Schiller2007,Derevianko2012}. However, locating the narrow optical clock transition remains a great challenge, in particular when there is no other known ancillary transition available. Ab-initio theory calculations typically predict transition energies with uncertainties at the percent level, corresponding to on the order of $10^{15}$ linewidths. Direct fluorescence spectroscopy in electron beam ion traps is limited to excited states with \textcolor{black}{lifetimes} of milliseconds \cite{lopez-urrutia_visible_2008}. Rydberg-Ritz combinations of such transitions can in some cases help narrow down the energy of the narrow clock transition \cite{bekker_detection_2019}. In general, the transition frequency needs to be found using laser spectroscopy techniques \cite{Chen2024}. This typically involves scanning a large bandwidth of possible transition frequencies with a narrow laser, which could result in months or even years of scanning before finding the transition. 

Such a ``needle in a haystack" problem is generic to quantum sensing: issues of similar nature arise in the context of dark matter search, identifying nuclear transitions and electronic transitions in superheavy elements or molecular ions \cite{Backes2021,Tiedau2024,Peik2005,wolf_non-destructive_2016}. Speeding up the spectroscopy while maintaining \textcolor{black}{a} high level of detection confidence often presents the main challenge. A well-known way to substantially enhance the detection efficiency is to employ squeezing in the detection protocol \cite{Schulte2018,zheng2016acceleratingdarkmatteraxionsearches,malia_free_2020,mao_quantum-enhanced_2023}. However, using squeezed or other engineered quantum states typically makes the protocol more fragile due to \textcolor{black}{the} inherent susceptibility of these states to background noise, heating and SPAM (State Preparation and Measurement) errors. Understanding the nontrivial interplay between these effects is thus crucial to \textcolor{black}{benefit} from squeezing and to reduce the \textcolor{black}{search} times in a realistic experimental environment.
\begin{figure}[t]
\centering
\centering
\includegraphics[width=.45\textwidth]{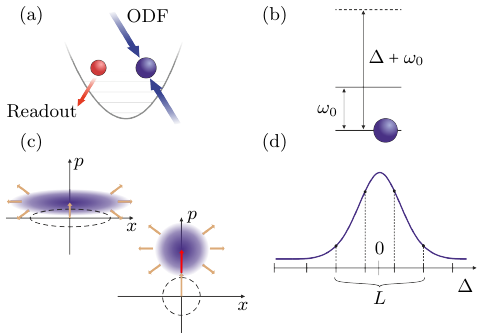}

\caption{\justifying Quantum logic spectroscopy protocol. (a) A spectroscopy ion with the sought-for electronic transition (blue, right) is trapped together with a logic ion (red, left), forming an ion crystal. (b) The unknown transition frequency $\omega_0$ is probed with an ODF interaction, parametrized by the interrogation detuning $\Delta$. (c) When tuned close to the transition, the ODF produces motional displacement and diffusion (yellow arrows) in the \textcolor{black}{squeezed} motional wave packet of the ion crystal's normal mode\textcolor{black}{. The packet is then ``unsqueezed'', amplifying the displacement signal, which is then} detected using the logic ion. (d) The ODF frequency detuning $\Delta$ is scanned step-wise. To detect the presence of the transition from the displacement signal, hypothesis testing is performed on the data from $L$ neighbouring frequency points.}
\label{fig:introfig}
\end{figure}

In this work, we theoretically study a quantum logic protocol used to search for a narrow transition in a trapped ion and ways to enhance the search efficiency in the presence of realistic trap heating and SPAM errors. For this, we develop a statistical framework and formulate the search problem as a statistical hypothesis test. The test itself is a deterministic procedure that takes in the measurement data in the vicinity of the interrogated frequency and decides whether the transition is likely present or not. The following steps are used to find the narrow transition within the large bandwidth. First, the ion is probed using an optical dipole force with a fixed interrogation frequency (Fig.\ref{fig:introfig} (a),(b)). Next, the motional displacement \textcolor{black}{of the momentum-squeezed motional state} is measured (Fig.\ref{fig:introfig} (c)). Then, the interrogation frequency is changed step-wise and the procedure repeats. Finally, the statistical test is performed on the data extracted from multiple frequencies (Fig.\ref{fig:introfig} (d)). Interrogation time, frequency step and number of measurements, together with variable parameters of the statistical test \textcolor{black}{and the amount of initial momentum squeezing} determine the performance of the \textcolor{black}{transition detection}. Using this approach, we describe how to optimize these parameters for maximum \textcolor{black}{transition search} speed, so that the prescribed confidence level is obeyed. We show that the postprocessing of the measurement data using statistical tests can ease the requirements on the \textcolor{black}{necessary} number of interrogations as well as other parameters and thus substantially increase the \textcolor{black}{search} speed. Furthermore, we demonstrate that using squeezed states of motion is advantageous compared to the vacuum state \textcolor{black}{due to much faster displacement signal generation.} The increased susceptibility to noise is effectively mitigated by using correlated statistical tests. 

The present work is organized as follows. We start by formulating the transition search problem and describe the microscopic model of the \textcolor{black}{recoil} spectroscopy protocol \textcolor{black}{in Section \ref{sec:microscopic}}. \textcolor{black}{To better understand the effects of motional squeezing, we employ a simplified analytical phase space model and discuss it in \textcolor{black}{Section} \ref{sec:squeezing}. In Section \ref{sec:statistical} we follow with} a description of our statistical model and the used hypothesis test. \textcolor{black}{Next, in Section \ref{sec:search}} we combine \textcolor{black}{the statistical and microscopic} models to describe a narrow transition search in a realistic experimental scenario. We discuss the \textcolor{black}{combined} effects of noise, squeezing and correlated data post-processing. \textcolor{black}{Furthermore, we find that the optimal value of the squeezing parameter lies within the range estimated using the analytical model.} \textcolor{black}{Finally, we draw conclusions and give a broader outlook on the results in Section \ref{sec:conclusions}.}

\section{Microscopic model}
\label{sec:microscopic}
We consider a problem of identifying the unknown frequency of a narrow electronic transition of a trapped ion using a quantum logic spectroscopy-like protocol. The transition is considered narrow if the uncertainty in transition energy is much larger than the linewidth of the detection signal. This happens, for instance, in case of clock transitions in highly charged or molecular ions. In highly-charged ions, the uncertainty is typically on the order of terahertz while the power-broadened linewidth is kilohertz-level. To search for the unknown frequency, we consider the probing interaction in the form of an optical dipole force (ODF), first demonstrated in the context of molecular ions \cite{wolf_non-destructive_2016} and recently applied to highly charged ions \cite{Chen2024}. In this scenario, the interrogated ion, in the following referred to as spectroscopy ion, is trapped together with another species \textcolor{black}{featuring a} well-controllable electronic structure. Together, the two trapped ions form an ion crystal with shared bosonic modes of motion, which serve as a bus for information transfer between the two electronic subsystems (Fig.\ref{fig:introfig}, (a)).

The ODF is created by applying two counter-propagating laser fields onto the ion crystal, tuned to differ by exactly the motional mode frequency $\nu=\omega_1-\omega_2$. The dynamics of the spectroscopy ion in the harmonic trap is governed by the following Hamiltonian, written up to the first-order Lamb-Dicke expansion in the proper rotating frame:
\begin{align}
    H=-\frac{\Delta}{2}\sigma_z+\frac{\Omega}{2}\sigma_x+\eta\frac{\Omega}{2}(\sigma^+a+\sigma^-a^\dagger).
    \label{eq:Ham}
\end{align}
Here, $\Delta=\omega_0-\omega_1$ is the detuning of the first driving field with respect to the atomic transition, $\sigma_i$ is a Pauli matrix with $i=\{x,y,z\}$, $\sigma^{\pm}$ are the two-level raising/lowering operators, $\Omega$ is the Rabi frequency \textcolor{black}{(set equal for both beams)}, $a^\dagger$ is the creation operator for the motional mode and $\eta$ is the Lamb-Dicke parameter, which includes the spectroscopy ion's participation amplitude in the in-phase mode of motion. \textcolor{black}{Tuned near to the unknown transition frequency $\omega_0$, such a field configuration exerts an effective force on the ions, causing a motional displacement. If the interrogation time is sufficiently long, the displacement can be detected via a red sideband transition on the logic ion \cite{Wan2014}, conditioning the spin excitation of the logic ion on the motional state of the two-ion crystal. The logic ion thus only serves for the readout and can be omitted from the microscopic description.} The time evolution of the system is described by the following master equation:
\begin{align} 
    \dot{\rho}=-\i[H,\rho]+\frac{1}{\tau_d}\mathcal{D}[\sigma^-]\rho+\frac{1}{\tau_h}\Big(\mathcal{D}[a]+\mathcal{D}[a^\dag]\Big)\rho.
    \label{eq:master}
\end{align}
Here the Lindblad dissipators $\mathcal{D}[L]\rho=L\rho L^\dag -\frac{1}{2}(L^\dag L\rho+\rho L^\dag L)$ account for the incoherent dynamics of the system. This includes spontaneous decay of the electronic excited state and linear motional heating in the trap, with rates defined by characteristic times $\tau_d$ and $\tau_h$ respectively. We note that for systems with longer probe times effects of motional dephasing might need to be taken into account. 

\textcolor{black}{The extracted signal $P_0$ is the fraction of the atomic population that has left the initial motional state $\ket{\psi_0}$. The detection stage is thus described by a binary positive operator-valued measure (POVM) consisting of complementary operators acting in the motional subspace:
\begin{align}
    \textrm{POVM}=\{\dyad{\psi_0},\mathds{I}-\dyad{\psi_0}\}
    \label{eq:POVM}
\end{align}}
\textcolor{black}{The simplest accessible choice for the initial motional state $\ket{\psi_0}$ is the vacuum state. However, further in Section \ref{sec:squeezing} we will show that using a momentum-squeezed state can benefit the displacement detection efficiency.}

\begin{figure}[t] 
\includegraphics[width=.47\textwidth]{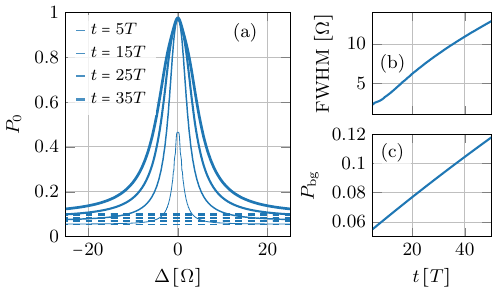}
\caption{\justifying (a) Excitation profile of the ODF interaction for different interrogation times \textcolor{black}{in the absence of squeezing.} The dashed horizontal lines in (a) correspond to the excitation due to the background noise in the absence of the transition.  (b) FWHM of the excitation profile. (c) Background noise in the absence of the excitation. \textcolor{black}{In all the subfigures, we assume the ODF with the parameters $\Omega=2\pi\times 5$~kHz and $\eta=0.1$. The decoherence times are fixed at $\tau_d=50T$ and $\tau_h=600T$.}}
\label{fig:lineshape}
\end{figure}

Performing interrogations at different values of the detuning $\Delta$ results in a spike-like fringe, \textcolor{black}{whose} profile and width depend on the interrogation time, see Fig.\ref{fig:lineshape}(a), solid lines.  Note that having larger $\tau_d$ will only \textcolor{black}{slightly} enhance the excitation profile, making the further results applicable for transitions with longer lifetimes. If the driving fields are tuned far away from the sought-for frequency $\omega_0$, the only contribution to the signal is \textcolor{black}{due to} the background noise Fig.\ref{fig:lineshape} (a), dashed lines, Fig.\ref{fig:lineshape}, (c). Since the possible \textcolor{black}{range of frequencies} is very large compared to both, the natural linewidth and the excitation width of the transition, locating it using ODF requires probing many interrogation frequencies step-by-step, until the transition signal can be confidently distinguished from the noise. The full width at half maximum of the signal grows with the probe time (Fig.\ref{fig:lineshape}, (b)) but so does the noise background (Fig.\ref{fig:lineshape}, (c)). This produces a non-trivial interplay: while a short interrogation time gives almost no background noise, the \textcolor{black}{displacement} signal is also very narrow and weak. A long interrogation time produces a larger signal, which is now hard to distinguish from the significant background noise\textcolor{black}{, see Section \ref{sec:squeezing} for more detail.}

Note that here we assume the probing interaction to be characterized by a single interrogation frequency, fixed during the interrogation time. The frequency is then varied step by step, until the transition is found. Although this procedure is quite natural to assume, there exist other approaches to transition searches, such as using adiabatic passages or frequency combs, \textcolor{black}{which have} advantages in certain experimental realizations \cite{Chen2024,Chou2020}. These techniques go beyond the scope of the present paper and are thus not discussed.

\section{Squeezing enhancement}
\label{sec:squeezing}
\begin{figure}[t]
\includegraphics[width=.45\textwidth]{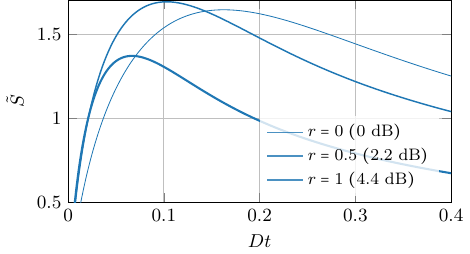}
\caption{\justifying \textcolor{black}{Logarithmic signal-to-noise ratio of the motional displacement signal for different amounts of initial momentum squeezing. The drift to diffusion ratio of the underlying Fokker-Planck equation \eqref{eq:FPE} is fixed at $\alpha/D = 10$ and generally depends on the unknown detuning parameter. Higher amounts of squeezing reduce the optimal interrogation time but may degrade the quality of the signal.}
}
\label{fig:squeezing_analytix}
\end{figure}

\textcolor{black}{In this section we discuss a simplified analytical model to show how motional squeezing can affect the detection efficiency. The master equation \eqref{eq:master} can be efficiently  reduced to a Fokker-Planck equation describing the dynamics of the motional Wigner function $W(x,p)$ in phase space \cite{carmichael_statistical_1999}:
\begin{align}
    \frac{\partial W}{\partial t}  = -\alpha\frac{\partial W}{\partial p}   + \frac{D}{2}\left(\frac{\partial^2 W}{\partial x^2} +\frac{\partial^2 W}{\partial p^2}    \right)
    \label{eq:FPE}
\end{align}
with $|\alpha|=\frac{\eta\Delta\Omega^2}{2\sqrt{2}(\Delta^2+\Omega^2)}$ and $D=1/\tau_h$ being respectively the drift and the diffusion coefficients of the Fokker-Planck equation. For details on the derivation, see Appendix \ref{app:FPE}. We now assume the motion to be prepared in a momentum-squeezed vacuum state $\ket{S(r)}=\e^{\frac{r}{2}(a^{\dagger2}-a^2)}\ket{0}$ with $\ket{0}$ being the vacuum state. We do not focus on a particular implementation of squeezing generation and will regard all possible imperfections of this process as SPAM-errors in the following sections. The real quantity $r$ is the squeezing parameter and translates to $-10\log_{10}\e^{-r}$ in dB units. The initial Wigner function of this state reads:
\begin{align}
    W_0(x,p)=\frac{1}{\pi}\exp(-e^{2r}p^2-e^{-2r}x^2).
\end{align}
The time evolution $W(x,p,t)$ of this state according to equation \eqref{eq:FPE} can be found analytically, and the time dynamics of the excitation signal $P_0$ of the logic ion is then calculated by evaluating the overlap between the initial and the time propagated phase space distributions:
\begin{align}
P_0(t)=1-2\pi\int \text{d}x\text{d}p\, W_0(x,p) W(x,p,t).
\label{eq:P0}
\end{align}
The quality of the excitation signal can be characterized by the logarithmic signal-to-noise ratio $\tilde{S}=\ln S\equiv\ln(P_0/P_\text{bg})$. Here, $S$ is the ratio of the excitation signal in the proximity to the transition and far detuned from it:
\begin{align}
    S= 1+\frac{1-\exp\left(-\frac{1}{2}\frac{\alpha^2t^2}{Dt+e^{-2r}}\right)}{\sqrt{1+D^2t^2+2Dt\cosh2r}-1}.
    \label{eq:SNR}
\end{align}
Expression \eqref{eq:SNR} determines the distinguishability of the displacement signal from the noise caused by heating. Near resonance, the detuning-dependent drift is stronger than the diffusion, $\alpha/D\gg1$, and high values of $\tilde{S}$ can be achieved given the probe time is small compared to the characteristic heating time of the trap: $Dt\ll 1$, see Fig.\ref{fig:squeezing_analytix}. For long probe times, the heating dominates the excitation signal and the signal-to-noise ratio decreases. The best detection efficiency is achieved at the intermediate probe time $t_{\text{opt}}$, which scales down with the squeezing parameter $t_{\text{opt}}\sim1/\alpha e^{r}$. This highlights an important feature: adding more squeezing to the protocol can produce a detectable displacement signal faster, thus speeding up the transition search. However, from Eq.\eqref{eq:SNR} and Fig.\ref{fig:squeezing_analytix} we see that the signal-to-noise ratio degrades if too much squeezing is used. Although squeezing reduces the shot noise in the displacement signal (seen by the growing numerator of the fraction in Eq.\eqref{eq:SNR}), it also significantly enhances the effective diffusion (denominator of the fraction in Eq.\eqref{eq:SNR}). For extreme amounts of squeezing ($r\gg1,\,Dt\gg e^{-2r}$), the diffusion dominates and the quality of the signal goes down exponentially with $r$: $S\sim e^{-r}/\sqrt{Dt}$. Strongly reducing the interrogation time, $Dt\ll e^{-2r}$ will also result in an unfavorable scaling: $S\sim Dt(\alpha/D)^2$.} \textcolor{black}{An optimal regime to use squeezing can be identified as $Dt\sim e^{-2r}$ and $(\alpha/D)^2Dt\gg1$. Given these relations, expression \eqref{eq:SNR} can be approximated by its upper bound $S\approx1+(\sqrt{1+Dt e^{2r}}-1)^{-1}$. For the transition search, we are mainly interested in reducing the probe time by maximizing squeezing, while also keeping the signal well-detectable, $\tilde{S}\approx1$. The maximal amount of squeezing obeying this constraint can be estimated as:
\begin{align}
    r^*=\frac{1}{2}\ln\left(\frac{\xi_0}{Dt}\right)
    \label{eq:squeezing_opt}
\end{align}
with $\xi_0=\left(1+\frac{1}{e-1}\right)^2-1$ and $e$ the Euler constant. The minimal value of the interrogation time $Dt$ is limited by the drift to diffusion ratio $\alpha/D$, which depends on the microscopic parameters of the experiment. In particular, it is proportional to the product of the sideband Rabi frequency and the heating rate of the trap, which generally needs to satisfy $\eta\Omega\tau_h\gg1$. Given the realistic parameters in Fig.\ref{fig:lineshape}, the drift to diffusion ratio spans the range $\alpha/D \sim 10\dots100$. This translates to a range of $10^{-3}\dots10^{-1}$ for the minimal interrogation time $Dt$ and the squeezing strength $6\dots16$ dB, evaluated using Eq.\eqref{eq:squeezing_opt}. Experimental conditions with significantly less noise and/or stronger signal can allow for using higher squeezing and shorter probe times.}

\textcolor{black}{The goal of the subsequent sections is to develop a more detailed statistical model for transition detection and apply it for balancing the competing effects of squeezing, heating and detection errors to maximize the transition search speed. A natural way to approach this problem is to employ the framework of hypothesis testing, a tool from statistical decision theory \cite{imagescience,Brubaker2017}.}

\section{Statistical model}
\label{sec:statistical}
To regard the transition as ``found'', we need to derive the corresponding criterion for the measurement data. For this, we do not limit ourselves to data from only one interrogated frequency. Instead, we analyze measurement data from several adjacent frequencies at once. If the step between the neighbouring points is not significantly larger than the excitation width, this helps to extract more information contained in the correlations between the neighbouring measurement frequencies. Note that the frequencies are still interrogated independently and correlations are only revealed in the postprocessing stage. 

From the statistical decision theory point of view, our task is to solve a \emph{classification} or a \emph{hypothesis testing} problem: we assume that the measured signal can only originate from two statistical models, with the first model describing the transition present within the target interval of frequencies and the other model corresponding to the background noise, which is the only contribution to the signal if the ODF is tuned far away from the transition. We thus define two hypotheses \cite{imagescience}:
\begin{align*}
    &H_1 \textrm{ – Transition is present (signal),}\\
    &H_2 \textrm{ – Transition is absent (background).}
\end{align*}
When deciding in favor of one or the other hypothesis, two distinct types of errors can be made: deciding in favor of the transition when none is present and deciding in favor of the background when there is in fact a transition within the analyzed spectral range. These are also known as errors of type I and II. The decision outcomes are summarized in Table \ref{tab:errors}.

\begin{table}[htbp]
\centering
\caption{Decisions and their outcomes}
\begin{tabular}{lllll}
\cline{1-3}
\multicolumn{1}{|l|}{}              & \multicolumn{1}{l|}{$H_2$: No transition}                                                         & \multicolumn{1}{l|}{$H_1$: Transition present}                                              &  &  \\ \cline{1-3}
\multicolumn{1}{|l|}{Decide for $H_2$} & \multicolumn{1}{l|}{True negative}                                                          & \multicolumn{1}{l|}{\begin{tabular}[c]{@{}l@{}}False negative\\ ``Miss"\end{tabular}} &  &  \\ \cline{1-3}
\multicolumn{1}{|l|}{Decide for $H_1$} & \multicolumn{1}{l|}{\begin{tabular}[c]{@{}l@{}}False positive\\ ``False alarm"\end{tabular}} & \multicolumn{1}{l|}{True positive}                                                   &  &  \\ \cline{1-3}
                                    &                                                                                             &                                                                                      &  & 
\end{tabular}
\label{tab:errors}
\end{table}
With the hypotheses formulated, we describe a deterministic procedure to accept or reject the hypothesis $H_1$ and assess the corresponding errors based on the input data. 

The measurement sequence at one fixed detuning point is modeled as a binary Bernoulli trial with $M$ independent measurements and the success probability is given by the excitation signal extracted from the microscopic model. If the transition is present, the Bernoulli parameter is $P_0(t,\Delta)$, if only the noise contributes it is given by $P_{\textrm{bg}}(t)$, which is the signal extracted from Eq.\eqref{eq:master} with $H=0$. Performing $M$ measurements at each of the $L$ neighbouring frequency points $\Delta_k$ produces the input data array $\mathbf{g}=(g_1, \dots, g_L)$, where each $g_k$ is the number of positive coin flip outcomes, i.e. an integer between zero and $M$. The conditional probability for the measured data to originate from one of the two hypotheses is given by:
\begin{align}
    &\mathrm{Pr}(\mathbf{g}|H_1)=\prod_{k=1}^{L}\binom{M}{g_k} P_0(t,\Delta_k)^{g_k} (1-P_0(t,\Delta_k))^{M-g_k},\\
    &\mathrm{Pr}(\mathbf{g}|H_2)=\prod_{k=1}^{L}\binom{M}{g_k} P_{\textrm{bg}}(t)^{g_k} (1-P_{\textrm{bg}}(t))^{M-g_k}.
\end{align}
To decide between $H_1$ and $H_2$ we need a \emph{test statistic} $\lambda(\mathbf{g})$, a real deterministic function of the data. The range of possible values of $\lambda(\mathbf{g})$ is partitioned into two regions, corresponding to deciding in favor of the hypothesis $H_1$ or $H_2$. In our case the range of possible values of the test statistic will be the whole real axis, split in two parts by a single tunable parameter $\Phi$. The regions are thus:
\begin{align*}
    &\mathbf{g}: \lambda(\mathbf{g})<\Phi \longrightarrow \textrm{decide for $H_1$},\\
    &\mathbf{g}: \lambda(\mathbf{g})\geq\Phi \longrightarrow \textrm{decide for $H_2$}.
\end{align*}
The choice of the test statistic $\lambda$ and the parameter $\Phi$ allows us to reach a desired balance between the two error probabilities. \textcolor{black}{In some search scenarios, the option to miss the transition is much more unfavorable than producing a false alarm, while in other cases, both error types need to be treated on equal footing.} For our binary classification problem, we choose a Neyman-Pearson test featuring a log-likelihood test statistic:
\begin{align}
    \lambda(\mathbf{g})=\ln\frac{\mathrm{Pr}(\mathbf{g}|H_2)}{\mathrm{Pr}(\mathbf{g}|H_1)}
    \label{eq:NPtest}
\end{align}
The miss rate ($\text{MR}$) and the false alarm ($\text{FA}$) probabilities are found as \textcolor{black}{a} sum over all the possible data arrays that generate the corresponding erroneous decision:
\begin{align}
    \text{MR}=\int_{\Phi}^{\infty}d\lambda \mathrm{Pr}(\lambda | H_1) = \sum_{\lambda(\mathbf{g})>\Phi} \mathrm{Pr} (\mathbf{g}|H_1),\label{eq:errors1}\\
    \text{FA}=\int_{-\infty}^{\Phi}d\lambda \mathrm{Pr}(\lambda | H_2) = \sum_{\lambda(\mathbf{g})<\Phi} \mathrm{Pr} (\mathbf{g}|H_2).
    \label{eq:errors2}
\end{align}

\begin{figure}[t]
\includegraphics[width=.47\textwidth]{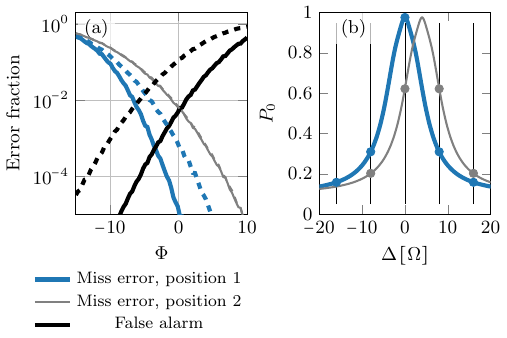}
\caption{\justifying (a) Decision errors for searching a narrow transition using a Neyman-Pearson hypothesis test. For a fixed parameter set \textcolor{black}{$L=5, M=8, t=35T, \Delta_s=8\Omega$}, the tunable test parameter $\Phi$ is striking the balance between the errors of two types (s. Tab. \ref{tab:errors}). Dashed lines represent noisy test with $10\%$ SPAM error. (b) The two considered positions of the signal lineshape with respect to the grid of interrogated frequencies. See text for details.
}
\label{Nptest}
\end{figure}

The statistical model of the signal requires the lineshape to have a fixed position with respect to the interrogated frequency grid. In reality, the offset with respect to the interrogation grid is unknown. To account for this uncertainty, in our model we assume two types of the lineshape position: one with the center of the signal aligned to the grid (Position 1) and one where it is shifted by half of the frequency step $\Delta_{\textrm{s}}/2$ (Position 2), see (b) panel of Fig.\ref{Nptest}. We conservatively estimate the miss rate of the test to be the greatest of the two position values: $\text{MR} = \max(\text{MR}_1, \text{MR}_2$). This approach is backed by simulations of statistical tests with a varying offset term, showing that these two cases cover the `worst possible' scenario of the signal positioning.

In our model, the output measurement data of a single interrogated frequency is a number between $0$ and $M$. Thus, we assume any SPAM-related error to manifest itself as a ``bit-flip" with a certain probability $\xi$. With this, any input data array $\mathbf{g}$ has a non-zero probability \textcolor{black}{$P_{\mathbf{g}\rightarrow\mathbf{g'}}$} to transform into any other input data array $\mathbf{g'}$. \textcolor{black}{To compute the net effect on the test error probabilities due to the noise, expressions (\ref{eq:errors1}, \ref{eq:errors2}) need to be supported with an additional weighted sum over all the possible transformations $\mathbf{g}\rightarrow\mathbf{g'}$. Such modification, together with computing all the transition probabilities would severely increase the numerical complexity of the error estimation. We note, however, that in a realistic scenario with $L>1$ and low SPAM error rate, only a small fraction of probabilities $P_{\mathbf{g}\rightarrow\mathbf{g'}}$ will differ from zero significantly. This observation allows us to efficiently employ a sampling technique to estimate the error rates affected by SPAM. Instead of the full computation, we sample the action of noise on a data vector for given $\xi$ and average the outcomes over many iterations.} Generally, the action of noise increases the error probabilities and reduces the performance of the test.

As an example, Fig.\ref{Nptest}(a) shows quantitatively the performance of a Neyman-Pearson test, in which $L=5$ adjacent frequency points are analyzed, each frequency being interrogated $M=8$ times with the interrogation time of $t=35T$ and the frequency step fixed at \textcolor{black}{$\Delta_{s}=8 \Omega$}. \textcolor{black}{The microscopic parameters used are the same as in Fig.\ref{fig:lineshape}}. The dashed curves show a noisy test with $\xi=10\%$ of SPAM errors. The two cases of the lineshape positioning are depicted in Fig.\ref{Nptest}(b). The value of $\Phi$ strikes the balance between the error probabilities of the test. As seen from panel (a), keeping both of the errors below $1\%$ in the noise-free case can be reached by setting $\Phi\approx0$. However, if $10\%$ noise is present, no value of $\Phi$ satisfies this constraint. In this case, a different parameter set needs to be taken to reach the $1\%$ confidence. \textcolor{black}{This could be achieved, for example, by using a finer detuning step $\Delta=5\Omega$. }

\textcolor{black}{One might ask a question of whether the described technique is still applicable in case when the laser noise cannot be neglected, which is a realistic scenario in search experiments \cite{Chen2024}. Generally, a laser-noise-induced dephasing reduces the peak value of the excitation profile and does not modify the background noise. For a broadband laser with noise on the level of the excitation Rabi frequency, the peak value reduces to about 50\% of the excitation. In this case, the method remains applicable, though the efficiency is reduced due to worse signal-to-noise ratio of the detection, which is confirmed by recent experiments in \cite{Chen2024,cheung_finding_2025}.}

\section{Narrow transition search}
\label{sec:search}
We can now combine both statistical and microscopical models to describe a realistic scenario of a narrow transition search using quantum logic spectroscopy and maximize its performance. \textcolor{black}{We fix a set of microscopic parameters corresponding to the experimental setup described in Ref.~\cite{King2022,Chen2024}. The Rabi frequency is fixed at $\Omega=2\pi\times 5$~kHz and defines a natural timescale of the problem $T=1/\Omega=0.2$~ms. The spontaneous decay rate is fixed at $\tau_d=50T=10$~ms and the trap heating rate is set to $\tau_h=600T$, which corresponds to heating of roughly $8.3$ phonons per second, the Lamb Dicke factor is $\eta=0.1$. The initial state of motion before the squeezing operation is assumed to be thermal, with $\bar{n}=0.05$ phonons on average. For the numerical solution of the master equation \eqref{eq:master}, the motional part of the Hilbert space is truncated after the first $30$ Fock states.}

The band scanning speed is found as the ratio of the frequency step $\Delta_s$ to the time needed for $M$ measurements:
\begin{align}
v=\frac{\Delta_s}{Mt}
\end{align}
Note that in a realistic scenario the interrogation time $t$ might include a fixed offset, which for squeezed states would also account for the overhead in state preparation and readout. For trapped ions, these operations are typically on the order of microseconds \cite{Burd2019}, much shorter than the interrogation time, and we thus omit this offset. When maximizing the search speed $v(M,t,\Delta_s)$, two non-linear constraints have to be met:
\begin{align}
    \text{MR}(L,M,t,\Delta_s)&<\epsilon_1,\label{eq:constraints} \\
    \text{FA}(L,M,t,\Delta_s)&<\epsilon_2.\nonumber
\end{align}
Here $\epsilon_1$ and $\epsilon_2$ are the two fixed confidence parameters determining the error rates of two kinds (see Table \ref{tab:errors}). While the function $v(M,t,\Delta_s)$ is easy to calculate and maximize over its domain, evaluating the error probabilities according to Eqs.(\ref{eq:errors1}, \ref{eq:errors2}) needs to be done numerically, which significantly adds computational complexity.

\begin{figure}[t]
\includegraphics[width=.52\textwidth]{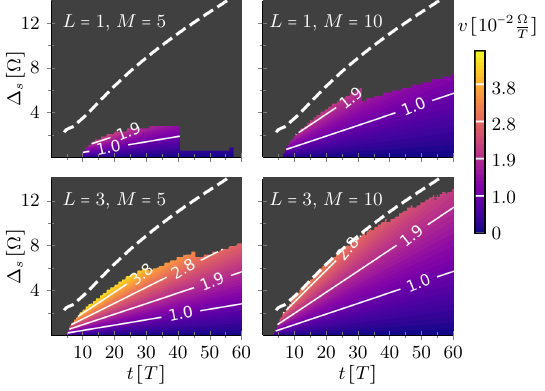}
\caption{\justifying Bandwidth search speed $v$ for different frequency steps $\Delta_s$ and interrogation times $t$. Each plot corresponds to a particular number of measurements $M$ and number of frequencies $L$ passed to the statistical test. Dashed white line is the FWHM of the signal lineshape.}
\label{4plots}
\end{figure}
\begin{figure}[t]
\includegraphics[width=.48\textwidth]{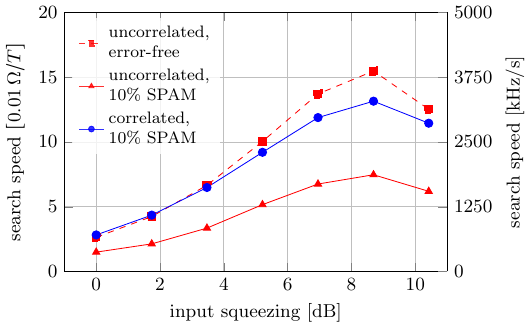}
\caption{\justifying Optimal bandwidth \textcolor{black}{search} speed versus the amount of squeezing used in the detection protocol. Hypothesis test analyzes data from single ($L=1$, red curves) or multiple ($L=3$, blue curve) interrogated frequencies. The dashed line corresponds to no SPAM errors, for the solid line $10\%$ chance of a bit-flip error in the data is assumed.}
\label{fig:sq_result}
\end{figure}
Let us first demonstrate how analyzing correlated data from multiple adjacent frequency bins ($L>1$) is beneficial opposed to the case of single-bin analysis ($L=1$), \textcolor{black}{which performance is comparable to standard signal-to-noise ratio techniques}. For this, we consider a simple case of a transition search using vacuum initial state in the absence of SPAM errors. The confidence parameters are set to $\epsilon_1=\epsilon_2=1\%$. Fig.\ref{4plots} shows the \textcolor{black}{search} speed $v$ as a function of $\Delta_s$ and $t$ for different $M$ and $L$. Masked gray region of the plot marks the parameter values that cannot satisfy the error constraints Eqs.\eqref{eq:constraints}. The white dashed line is the FWHM of the signal (see Fig.\ref{fig:lineshape}). Comparing the two rows, one sees a significant improvement in the maximal achievable search speed, if $L=3$ frequency bins are analyzed. As more data is given to the test, the error probabilities \textcolor{black}{decrease}, unlocking more advantageous regions in the $(\Delta_s,t)$ plane. Generally, increasing $L$ will be beneficial as long as the involved frequencies still fall in a proximity of the \textcolor{black}{signal} lineshape. The two columns of Fig.\ref{4plots} demonstrate that increasing the number of measurements unlocks even greater regions of the plane, but it also takes more time and hence might ultimately reduce the search speed. Thus, an optimum needs to be found in the parameter space $(M,t,\Delta_s)$ to maximize the \textcolor{black}{frequency search} speed.

We now consider a squeezed initial state and the narrow transition search using the corresponding squeezed measurement. In this case, SPAM errors can contribute significantly and will be modeled at $\xi=10\%$ level with the Monte-Carlo method, as described \textcolor{black}{at} the end of previous section. The squeezing parameter $r$ is varied up to about $10$ dB, which is a realistic amount of squeezing for current experiments \cite{lo_spin-motion_2015,Burd2019,Drechsler2020}. For each value of the squeezing parameter, we fix the confidence level at $1\%$ for both error types and search for the maximum value of the \textcolor{black}{frequency search} speed. The results are summarized in Fig.\ref{fig:sq_result}, \textcolor{black}{which show} the curves for correlated ($L=3$) and uncorrelated ($L=1$) statistical tests with the dashed curve representing the error-free case. For any statistical test used, in both noise-free and noisy measurements, we observe a significant boost in bandwidth \textcolor{black}{search} speed offered by squeezing, with the \textcolor{black}{greatest} advantage achieved at around \textcolor{black}{$8-9$} dB for the experimental parameters considered\textcolor{black}{, which falls well within the $6-16$ dB range found analytically in Section \ref{sec:squeezing}}. Higher amounts of squeezing demand \textcolor{black}{a} greater number of measurements and are thus slower. This happens due to the increased sensitivity of the highly-squeezed states to the background heating noise.

As expected, the action of SPAM errors reduces the performance of the statistical test and slows down the narrow transition search by roughly a factor of two for any amount of squeezing considered. As well as for the vacuum readout, the correlated test enables faster transition search. We find that the two competing effects produce comparable change in the optimal \textcolor{black}{search} speed: the noiseless curve for $L=1$ lies close to the $L=3$ curve with 10\% SPAM errors. Thus, the action of SPAM errors at this level can effectively be mitigated by correlated statistical tests. Together with realistic \textcolor{black}{$8-9$} dB of squeezing, this improves the \textcolor{black}{search} speed by roughly an order of magnitude, when compared to the single-frequency test with the vacuum initial state (no squeezing). For the experimental parameters considered, this means an increase of the bandwidth \textcolor{black}{search} speed from \textcolor{black}{$374$} kHz/s to \textcolor{black}{$3.3$} MHz/s. The typical level of uncertainty of the ab-initio atomic structure calculations for optical transitions in highly charged ions is of the THz order. For such a bandwidth, the presented technique reduces the transition search time from \textcolor{black}{$31$} to \textcolor{black}{$3.5$} days, while staying at the 99\% confidence level. \textcolor{black}{With additional tests we found that the optimal value of the squeezing parameter and qualitative shape of the curves in Fig.\ref{fig:sq_result} remain largely independent \textcolor{black}{of} the heating rate of the trap.}

\section{Conclusions and outlook}
\label{sec:conclusions}
In this study we showed how combining correlated statistical hypothesis testing and motional squeezing can lead to a significant metrological gain in quantum logic spectroscopy. In a realistic case study motivated by recent experiments we demonstrated how a search for \textcolor{black}{a} narrow electronic transition in a trapped ion can be sped up by \textcolor{black}{about} an order of magnitude by using \textcolor{black}{a} reasonable amount of squeezing and correlated statistical tests, compared to the usual vacuum interrogation and uncorrelated data analysis. This is a significant reduction \textcolor{black}{in} experimental overhead, potentially reducing the required experiment runtime to find the transition from several months to about one week. 
Similar machinery can be applied to various physical problems that involve scanning large bandwidths of frequencies in search \textcolor{black}{of} signals of predictable shape. The mathematical apparatus of statistical decision theory is shown to be a powerful tool and it is particularly useful when multiple effects need to be taken into account, such as SPAM-errors, intrinsic noise coming from heating and sensitivity gain from engineered quantum states. Though it has not been explored extensively in this paper, the presented approach also offers flexibility to balance the different error types.  For instance, in ``needle in a haystack'' problems it might \textcolor{black}{prove} itself beneficial to put a tighter confidence constraint on false negative error fraction while relaxing the bound for the false positive detections. For transition searches, this would mean even higher search speed at the cost of more frequent false alarms. Further studies on enhancing searches for narrow transitions might include use of more exotic engineered quantum states as well as different interrogation techniques, i.e. utilizing frequency combs or quantum information processing.

\textcolor{black}{Furthermore, we used an analytical model based on the Fokker-Planck equation for the phase space distribution to estimate the amount of squeezing required for the search speed enhancement. The value found by the full numerical optimization of the line search problem using statistical tests fits well within the theoretically estimated range.}

The mathematical apparatus of statistical decision theory can extend far beyond what has been presented in this paper. Particular extensions might include use of different statistical tests, more advanced probability and noise models and more efficient computational methods. To ease the implementation overhead, we provide a code repository with the discussed hypothesis testing implemented in Python and thoroughly documented \cite{CODEREPO}.

\section*{Acknowledgements}
The project was supported by the Physikalisch-Technische Bundesanstalt, the Max--Planck--Riken--PTB--Center for Time, Constants and Fundamental Symmetries, and the Deutsche Forschungsgemeinschaft (DFG, German Research Foundation) through SCHM2678/5-2, the collaborative research center SFB 1227 DQ-mat (Project-ID 274200144), from BMBF through the project ‘ATIQ’, and under Germany’s Excellence Strategy -- EXC-2123 QuantumFrontiers -- 390837967. This project has received funding from the European Research Council (ERC) under the European Union’s Horizon 2020 research and innovation program (grant agreement No 101019987).

%

\appendix
\section{Fokker-Planck equation}
\label{app:FPE}

\textcolor{black}{
Our goal is to find an effective description for the dynamics of Eq.\eqref{eq:master} in terms of the Wigner function of the motional mode. We will neglect the spin decay in this treatment, since for realistic probe times it only has a weak effect on the displacement signal dynamics. Thus, the incoherent part of the dynamics only involves motional heating, which has no effect on the spin degree of freedom. With this assumption, let us first rewrite Eq.\eqref{eq:master} in terms of the position and momentum variables:
\begin{align}
    \dot{\rho}=-\i[H_0+\eta\tilde{\Omega}\left(x\sigma_x-p\sigma_y\right),\rho]+\frac{1}{\tau_h}\left(\mathcal{D}[x]+\mathcal{D}[p]\right)\rho,
\end{align}
where we defined $H_0=-\frac{\Delta}{2}\sigma_z+\frac{\Omega}{2}\sigma_x$ and $\tilde{\Omega}=\frac{\Omega}{2\sqrt{2}}$. Next, we do a Wigner-Weyl transformation for the bosonic mode, giving us the equation for the Wigner representation $w(x,p)$, which remains a density matrix for the spin:
\begin{align}
    \dot{w}=\mathcal{L}_0w+\mathcal{L}_{\text{int}}w+\mathcal{L}_{\text{heat}}w
\end{align}
with
\begin{align}
    \mathcal{L}_0w&=-\i[H_0,w],\\
    \mathcal{L}_{\text{int}}w&=-\i\eta\tilde{\Omega}\Big( (x+\frac{\i}{2}\partial_p)\sigma_xw-(x-\frac{\i}{2}\partial_p)w\sigma_x,  \\
    &-(p+\frac{\i}{2}\partial_x)\sigma_yw+(p+\frac{\i}{2}\partial_x)w\sigma_y\Big),\\
    \mathcal{L}_{\text{h}}w&=\frac{1}{2\tau_h}\left(\partial^2_x+\partial^2_p\right)w.
\end{align}
For the transformation, the standard set of the replacement rules is used \cite{gardiner_quantum_2010}:
\begin{align}
    x\rho&\rightarrow\left(x+\frac{\i}{2}\partial_p\right)w,\\
    \rho x&\rightarrow\left(x-\frac{\i}{2}\partial_p\right)w,\\
    p\rho&\rightarrow\left(p-\frac{\i}{2}\partial_x\right)w,\\
    \rho p&\rightarrow\left(x+\frac{\i}{2}\partial_x\right)w.
\end{align}
We now do an approximation by tracing out the spin degrees of freedom and introduce the Wigner function $W(x,p)=\Tr_s[w]$. The equation of motion takes the form of a Fokker-Planck equation:
\begin{align}
    \dot{W}=\left\{\eta\tilde{\Omega}\left(\partial_p\langle\sigma_x\rangle+\partial_x\langle \sigma_y\rangle\right)+\frac{1}{2\tau_h}\left(\partial^2_x+\partial^2_p\right)\right\}W.
    \label{eq:FPE0}
\end{align}
The time dynamics of the spin degrees of freedom is found by solving the optical Bloch equations with the Hamiltonian $H_0$ for an initial spin ground state. The relevant expectation values are:
    \begin{align}
        \langle\sigma_x\rangle&=\frac{\Delta\Omega}{\Omega'^2}\left(\cos(\Omega't)-1\right),\\
        \langle\sigma_y\rangle&=\frac{\Omega}{\Omega'}\sin(\Omega't)
    \end{align}
with $\Omega'^2=\Omega^2+\Delta^2$. On the timescales $t\gg1/\Omega'$, $\langle\sigma_y\rangle$ averages to zero, thus the Fokker-Planck equation \eqref{eq:FPE0} has effectively no drift dynamics in the coordinate quadrature. For the momentum quadrature, the drift coefficient describes the net force exerted on the ions:
    \begin{align}
        \alpha=-\frac{\eta}{2\sqrt{2}}\frac{\Omega^2}{\Omega'^2}\Delta.
    \end{align}
One might notice that on resonance, $\Delta=0$, the average exerted force is zero. In fact, this is not surprising, since on resonance the ODF Hamiltonian \eqref{eq:Ham} creates a Schrödinger cat-like state with the two constituents being pulled in opposite directions. To see this, we can look at the ODF interaction in the rotating frame with respect to $H_0$.  The time-dependent terms of the resulting expression can be dropped in a rotating wave approximation, provided $\eta\ll1$:
    \begin{align}
    \tilde{H}=\eta\tilde{\Omega}\left(x\sigma_x-p(\sigma_y\cos\Omega t - \sigma_z\sin\Omega t)\right)\approx\eta\tilde{\Omega}x\sigma_x.
    \end{align}
The resulting effective Hamiltonian $\tilde{H}$ describes a spin-dependent force that has a different sign for the two eigenstates of the $\sigma_x=\dyad{+}_x-\dyad{-}_x$ operator. Since the initial ground state is an equal superposition of $\sigma_x$ eigenstates, $\ket{g}=\frac{1}{\sqrt{2}}(\ket{+}_x-\ket{-}_x)$, the time evolution splits the motional Wigner function into two constituents with opposite momenta. Although this formally results in a suppressed drift coefficient in the case of $|\Delta|\ll\Omega$, a strong excitation signal will still be obtained on the logic ion, since the motion gets excited and the system quickly leaves the initial state. Outside of this practically unlikely regime, the displacement signal can be found by directly solving the Fokker-Planck equation.
}

\textcolor{black}{
We can now write Eq.\eqref{eq:FPE0} in a canonical form (see Eq.\eqref{eq:FPE}) and time propagate it using the corresponding Green's function:
\begin{align}
    W(x,p,t)=\int \text{d}x'\text{d}p'\, W_0(x',p') G(x,x',p,p',t),
\end{align}
where $G(x,x',p,p',t)=\frac{1}{2\pi D t} \exp \left(-\frac{(x-x')^2}{2Dt}-\frac{(p-p'-\alpha t)^2}{2Dt}\right)$. The excitation signal is then found by directly evaluating the expression \eqref{eq:P0}:
\begin{align}
    P_0=1-\frac{\exp\left(-\frac{1}{2}\frac{\alpha^2t^2}{e^{-2r}+Dt}\right)}{\sqrt{(1+e^{-2r}Dt)(1+e^{2r}Dt)}}.
\end{align}
}

\end{document}